\documentclass[aps,amsfonts,nofootinbib,twocolumn]{revtex4}
\usepackage{epsfig}
\usepackage{graphicx}
\usepackage{amsmath}
\begin{document}
\newcommand{\bb}{\begin{equation}}
\newcommand{\ee}{\end{equation}}
\newcommand{\eqb}{\begin{eqnarray}}
\newcommand{\eqf}{\end{eqnarray}}
\frenchspacing
\newcommand{\hs}{/\kern-.52em h}
\newcommand{\D}{{\cal D}}
\newcommand{\f}{F^{\mu\nu}}
\newcommand{\DB}{\delta_{BRST}}
\newcommand{\QB}{Q_{BRST}}
\newcommand{\GG}{\Gamma^{2n}}
\newcommand{\GGG}{\Gamma^{*2(n-m)}}
\newcommand{\med}{\D \bar \psi \D \psi}
\newcommand{\medp}{\D \bar \psi' \D \psi'}
\newcommand{\id}{i\kern.06em\hbox{\raise.25ex\hbox{$/$}\kern-.60em$\partial$}}
\newcommand{\as}{/\kern-.68em A}
\newcommand{\Ds}{/\kern-.69em D}
\newcommand{\vs}{\varphi^{0}_n}
\newcommand{\ks}{/\kern-.67em k}
\newcommand{\Ps}{/\kern-.65em p}
\newcommand{\uh} {\frac{1}{\hbar}}
\newcommand{\BBs}{\!\not\!\! B}
\newcommand{\rD}{\!\not\!\! D}
\newcommand{\bs}{/\kern-.52em b}
\newcommand{\qs}{/\kern-.52em s}
\newcommand{\dv}{\!d^3\!x\,}
\newcommand{\Z}{{\cal Z}}
\frenchspacing

\preprint{}

\title{ Chiral Anomaly Beyond  Lorentz Invariance}
\author{Paola Arias}
\email{paola.arias@gmail.com} \affiliation{Departamento de F{\'\i}sica,
Universidad de Santiago de Chile, Casilla 307, Santiago, Chile}
\author{H. Falomir}
\email{falomir@fisica.unlp.edu.ar} \affiliation{IFLP - Departamento
de F{\'\i}sica, Facultad de Ciencias Exactas, Universidad Nacional de la
Plata, C.C. 67, (1900) La Plata, Argentina}
\author{J. Gamboa}
\email{jgamboa@usach.cl} \affiliation{Departamento de F{\'\i}sica,
Universidad de Santiago de Chile, Casilla 307, Santiago, Chile }
\author{F. Mendez}
\email{fmendez@usach.cl} \affiliation{Departamento de F{\'\i}sica,
Universidad de Santiago de Chile, Casilla 307, Santiago, Chile }
\author{F. A. Schaposnik }\thanks{Associated with CICBA}
\email{fidel@fisica.unlp.edu.ar}\affiliation{IFLP - Departamento de
F{\'\i}sica, Facultad de Ciencias Exactas, Universidad Nacional de la
Plata, C.C. 67, (1900) La Plata, Argentina\\
 CEFIMAS-SCA, Ave. Santa Fe 1145, C1059ABF, Buenos Aires, Argentina}

\begin{abstract}
The chiral anomaly in the context of an extended standard model with
Lorentz invariance violation  is studied. Taking into account bounds
from measurements of the speed of light, we argue that the chiral
anomaly and its consequences are general results valid even beyond
the relativistic symmetry.
\end{abstract}

\maketitle

\section{Introduction}
Lorentz invariance is a cornerstone of relativistic Quantum Field
Theory \cite{weinberg}. However, in the last years  many authors
have argued that at very high energy some symmetries, such as
Lorentz and CPT, could be broken and, therefore, new scenarios and
physical processes could take place
\cite{CPT,kost,jackiw,nos,jackiw1,bertolami}.

Notice that Minkowski space and its isometries, the Lorentz
transformations, should arise from a low energy solution of string
theory. Therefore, it is a legitimate question to ask about a
possible relic of this origin in a QFT at high energies.

However, even if a foremost invariance as the Lorentz one could be
broken, one should expect that some important features and
properties of quantum field theory are preserved, as well as the
stability of some related phenomena.

In the context of a relativistic QFT with gauge fields and fermions,
the chiral anomaly is related, as it is well known,  to a
topological object and is thus independent of the energy scale. This
suggests that its form could be preserved even if Lorentz and CPT
symmetries were broken and, in particular,  the $\pi^0 \rightarrow 2
\gamma$ decay as derived from the chiral anomaly should be
unaffected at any energy scale. So, it is worthwhile to consider
this possibility in the framework of a quantum field theory model
with an explicit Lorentz and/or CPT symmetry breaking.

The purpose of this note is to analyze the fate of the chiral
anomaly in a model with an explicit Lorentz symmetry breaking and to
explore, in connection with this, whether its role in connection
with the $\pi^0 \to 2\gamma$ process is affected.

\section{A Lorentz symmetry violating model for fermions}

Let us start by considering the $d=4$ dimensional Lagrangian
\cite{mewes}
\begin{equation} {\cal L} =  {\bar \psi_q}\, \Gamma^\mu D_\mu (A)\psi_q \; ,
\label{1} \end{equation}
where $A_\mu = A_\mu^a T_a$ ($\mu = 0,1,2,3 \, ; \; a=1,2,\ldots,
{\rm dim}{\cal G}$) are gauge fields taking values in the Lie
algebra of some gauge group $G$ with generators $T_a$. Dirac
fermions $\psi_q$  are taken in the fundamental representation of
$G$ and $D_\mu$ is the usual covariant derivative. Concerning
matrices $\Gamma^\mu$, they can in general take the  form
\begin{equation}
 \Gamma^\mu = \gamma^\mu + \Gamma^\mu_{LV}
+ \Gamma^\mu_{CPTV}\,. \label{2} \end{equation}
Here $\gamma^\mu$ are the usual Dirac matrices while
$\Gamma^\mu_{LV}$ and $\Gamma^\mu_{CPTV}$ are matrices which
introduce violation of Lorentz and Lorentz-CPT symmetries
respectively. They are defined as
\eqb \Gamma^\mu_{LV} &=& c^\mu_{\ \nu}\gamma^\nu + d^\mu_{\ \nu}
\gamma^\nu \gamma_5, \nonumber
\\
\Gamma^\mu_{CPTV}&=& e^\mu + f^\mu \gamma_5+ g^{\mu \nu
\lambda}\sigma_ {\nu \lambda}, \label{te1} \eqf
where $c^\mu_{\ \nu},d^\mu_{\ \nu},e^\mu,f^\mu$ and $g^{\mu
\nu\lambda}$  are real constants to be, in principle,
phenomenologically determined.

 But, if the theory is required to be invariant under
(global) chiral transformations of the fermionic field at the
classical level, then the CPT violating terms must be removed
since
\begin{equation}\label{no-CPTV}
    \left\{\gamma_{5} , \Gamma^{\mu}_{CPTV}\right\} \neq 0\,.
\end{equation}
 So, we will take $e^\mu=0$, $f^\mu=0$ and $g^{\mu
\nu\lambda}=0$.

\medskip

 On the other hand, although the $\Gamma^\mu$ matrices
formally play the role of Dirac matrices, they do not satisfy
in principle the standard Clifford algebra.  Indeed, if we write
\begin{equation}\label{G1}
    \Gamma^{\mu}=\omega^{\mu}_{\ \nu} \gamma^{\nu}
    + d^{\mu}_{\ \nu} \gamma^{\nu} \gamma_{5}\,,
\end{equation}
where
\begin{equation}\label{omega}
    \omega^{\mu}_{\ \nu} = \delta^{\mu}_{\ \nu} + c^{\mu}_{\ \nu}\,,
\end{equation}
it is straightforward to get
\begin{equation}\label{G2}
    \begin{array}{c}
      \left\{  \Gamma^{\mu} ,  \Gamma^{\nu} \right\}=
    \left( \omega^{\mu}_{\ \alpha} \omega^{\nu}_{\ \beta }
    - d^{\mu}_{\ \alpha} d^{\nu}_{\ \beta}\right) \left\{  \gamma^{\alpha} ,  \gamma^{\beta} \right\} +
    \\ \\
      +  \left( \omega^{\mu}_{\ \alpha} d^{\nu}_{\ \beta }
    - d^{\mu}_{\ \alpha} \omega^{\nu}_{\ \beta}\right) \left[  \gamma^{\alpha} ,  \gamma^{\beta} \right]
    \gamma_{5} \,.
    \end{array}
\end{equation}
Then, if (in order to get a Clifford algebra for the
$\Gamma$-matrices) we demand the last term in the right hand side
not to be present, we must impose that
\begin{equation}\label{G3}
    \left( \omega^{\mu}_{\ \alpha} d^{\nu}_{\ \beta }
    - d^{\mu}_{\ \alpha} \omega^{\nu}_{\ \beta}\right)
    -\left( \omega^{\mu}_{\ \beta} d^{\nu}_{\ \alpha }
    - d^{\mu}_{\ \beta} \omega^{\nu}_{\ \alpha}\right) =0\,.
\end{equation}
Multiplying by $\left(\omega^{-1}\right)^{\alpha}_{\ \mu}$ (notice
that $\omega^{\mu}_{\ \alpha}$ is invertible, since we are looking
for small LIV) we get
\begin{equation}\label{G4}
   4 d^{\nu}_{\ \beta } =
    \left[ \left(\omega^{-1}\right)^{\alpha}_{\ \mu} d^{\mu}_{\ \alpha} \right] \omega^{\nu}_{\ \beta}
    \,,
\end{equation}
whose general solution is
\begin{equation}\label{G5}
    d^{\nu}_{\ \beta } = Q \  \omega^{\nu}_{\ \beta}
\end{equation}
with $Q$ a constant.

 Consequently, we restrict our attention to this
\emph{minimal} Lorentz invariance violation, preserving chiral
symmetry at the classical level and the form of the Clifford
algebra, and take
\begin{equation} \Gamma^\mu = \omega^{\mu}_{\ \nu}\gamma^\nu \left(\mathbf{1}_4
+Q\, \gamma_{5} \right) \,. \label{true} \end{equation}

  Notice that
\begin{equation}\label{G6}
          \left\{  \Gamma^{\mu} ,  \Gamma^{\nu} \right\}_{\pm}=
    \left( 1  -Q^{2} \right)
    \omega^{\mu}_{\ \alpha} \omega^{\nu}_{\ \beta }
    \left\{  \gamma^{\alpha} ,  \gamma^{\beta} \right\}_{\pm} \,,
\end{equation}
where we have taken $Q^2 \ll 1$ since we shall consider small
deviations from Lorentz invariance.

 Therefore, the chosen set of $\Gamma$-matrices does fulfill
the relations
\eqb \{ \Gamma^\mu, \Gamma^\nu\} &=& 2 M^{\mu \nu}\, \mathbf{1}_4,
\label{cli}
\\
\{\Gamma^\mu, \gamma_5\} &=&0\,,\label{07} \eqf
where $M^{\mu \nu}$ is a metric like object defined as
\begin{equation} M^{\mu \nu} = \Omega^{\mu}_{\ \alpha} \Omega^{\nu}_{\ \beta} \,
\eta^{\alpha \beta}\,, \label{metric} \end{equation} where
$\Omega^{\mu}_{\ \alpha} = \omega^{\mu}_{\ \alpha}\, \sqrt{ 1
-Q^{2}} $ and $\eta^{\alpha \beta}$ is the standard metric in
Minkowski space.

\medskip
  Notice  that, since we are only interested in Lorentz
violation effects in the fermion sector, we are also omitting a
possible LIV term in the photon sector given by  \cite{ko} $(
\kappa_F)_{k\lambda\mu\nu}F^{k\lambda}F^{\mu\nu}$.

\medskip

We shall present in the next section a derivation of the chiral
anomaly for a quantum field theory in which the  Fermi fields
 dynamics is governed   by  the Lorentz violating
 fermion  Lagrangian in Eq.\ (\ref{1}), testing whether
 the index theorem is still valid.   We shall then see that neither the
anomaly (Eq.\ (\ref{021})) nor the the index theorem (Eq.\
(\ref{031})) are affected by the minimal Lorentz symmetry violation
introduced in $\Gamma^{\mu}$. Only the axial current
$\mathcal{J}^{\mu}_{5}$ is changed into
\begin{equation}\label{j5}
    \mathcal{J}^{\mu}_{5}= \omega^{\mu}_{\ \nu}
    \left( j^{\nu}_{5}+Q  j^{\nu} \right)\,,
\end{equation}
where $j^{\nu}_{5}$ is the axial vector current arising in the
ordinary Lorentz invariant case, $j^{\mu}_{5}= \bar{\psi_q}
\gamma^\mu\gamma_5 \psi_q$, and $j^{\mu}= \bar{\psi_q} \gamma^\mu
\psi_q$ is the vector current.

Before doing this,  we shall   discuss the situation from a more
phenomenological point of view. To start up, let us assume that
$c^{\mu}_{\ \nu}$ has only one non-zero component, namely
$c^{0}_{\ 0}=\kappa$.   With this,  rotational invariance is
preserved and then
\begin{equation}\label{100}
    \begin{array}{c}
      \Gamma^{0}=\left(1+c^{0}_{\ 0}\right) \gamma^{0}\left(\mathbf{1}_4 +
      Q\, \gamma_{5} \right)
       \,, \\ \\
      \quad \Gamma^{i}=\gamma^{i} \left(\mathbf{1}_4 +Q\, \gamma_{5} \right) \,,
    \end{array}
\end{equation}
and
\begin{equation}\label{101}
   \left( M^{\mu \nu} \right) =
   \left( 1 - Q^2\right) {\rm diag}\left(
   (1+\kappa)^2,-1,-1,-1\right)\,.
\end{equation}
  When replaced in the modified Dirac equation following from
Lagrangian  (\ref{1}), this leads to the (free) dispersion relation
\begin{equation}\label{102}
    \begin{array}{c}
      \Gamma^{\mu}  \Gamma^{\nu} \,p_{\mu} p_{\nu}
    = M^{\mu \nu}\,p_{\mu} p_{\nu} = \\ \\
      =\left( 1 - Q^2\right)
    \left\{\left(1+\kappa\right)^2 {p_{0}}^{2}-{\textbf{p}}^{2}\right\}=0\,,
    \end{array}
\end{equation}
 with $p_0=E/c$, where $c$ is the standard value of the velocity of light
used here to set the length scale.
 Eq.\ (\ref{102}) implies that \emph{massless} fermions $\psi_q$ move
 with velocity $v_q$
given by
\begin{equation}\label{103}
    v_q=\frac{c}{1+\kappa}
\end{equation}
(with no dependence on the parameter $Q$).


\begin{figure}[h]
\begin{center}
  \includegraphics[width=.4\textwidth]{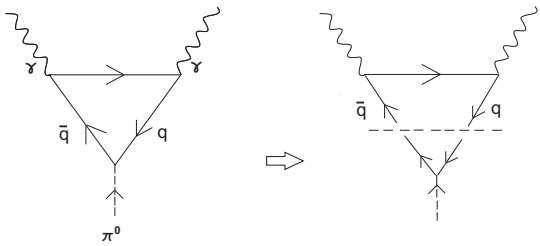}\\
  \caption{Feynman diagram for pion decay.}\label{fig}
   \end{center}
\end{figure}


Eq.(\ref {103}) of course implies new physics, in the vein of
\cite{cole1}-\cite{cole} (see \cite{GG} for a complete list of
references). Let us advance, however, that the results in the next
section show that the minimal Lorentz invariance violation
introduced here  does not induce modifications in the form of the
chiral anomaly. Nevertheless, it could imply modifications in the
calculation of physical observables.

\medskip

 We shall now contrast our results with experiment
by considering, in a Lorentz violating invariance framework, the
celebrated $\pi^0 \rightarrow 2 \gamma$ process
\cite{adler,jackiw2}, directly related to chiral anomaly. In
particular,  let us analyze whether a departure from the usual
chiral anomaly takes place by assuming that a formula similar to
(\ref{103}) (obtained for the case of massless fermions) also holds
for photons.
This can be justified as follows. In the present Lorentz invariance
violating framework, one can relate the velocity of the photons with
that of fermions in a given vertex through the formula
\begin{equation}
    c_{ph} = n_0 v_q \label{photon}
\end{equation}
where we have defined
\begin{equation}
    n_0 = \frac{E_{ph}}{E_q}
\end{equation}
with $E_{ph}$ and $E_q$ the energies of photons and fermions
respectively

Formula (\ref{photon}) can be obtained by cutting the standard
triangle diagram (see fig. \ref{fig}) and using the 4-momentum
conservation law for the process $\pi^0 \rightarrow 2 \gamma$. Doing
this one has \begin{equation} \left( \frac{E_q}{v_q}, {\vec
P}\right) + \left( \frac{E_{\bar q}}{v_q}, -{\vec P}\right) = 2
\left( \frac{E_{ph}}{c_{ph}},{\vec 0}\right), \end{equation} and
therefore
\begin{equation} \frac{E_q + E_{\bar q}}{v_q} = 2
\frac{E_{ph}}{c_{ph}}. \end{equation}

Since   we are assuming that
CPT invariance is conserved, $E_q=E_{\bar q}$ and then Eq.\
(\ref{photon}) follows.

We shall consider the case in which the photon velocity $c_{ph}$
does not exceed that of the neutral pion $v_{\pi^0}$ (in the case
$c_{ph} > v_{\pi^0}$ the decay is kinematically forbidden
\cite{cole}). Also, interpreting   $n_0$ in Eq.\ (\ref{photon}) as a
refraction index, it is natural to take $n_0 \approx 1$ in empty
space. With all this, the off-mass-shell decay amplitude in the
chiral limit is


\eqb (T(\pi^0 \rightarrow 2 \gamma)_{LIV} &=& \frac{\alpha^2}{64
\pi^3} \left( \frac{m_\pi}{f_\pi}\right)^2 m_\pi c^6_{ph} \nonumber
\\
&=& \frac{\alpha^2}{64 \pi^3} \left(
\frac{m_\pi}{f_\pi}\right)^2   \frac{m_\pi c^6}{(1+ \kappa)^6}. \label{1f} \eqf

One can now estimate the ratio $T_R/T_{LIV}$, with $T_R$ the
relativistic rate of decay, using the bounds  for $\kappa$ discussed
in the literature. The comparison between the Lorentz violating
invariance scheme and the relativistic one yields to
\begin{equation} \frac{(T)_{R}}{(T)_{LIV}}-1\approx 6 \kappa. \label{esti}
\end{equation}
In order to estimate the right hand side in (\ref{esti}) let us
introduce $\delta c$ through the equation \begin{equation} \delta c
= c - c_{ph} \label{primera} \end{equation} where $c$ is the
standard value of light velocity. Different experimental and
phenomenological tests show the smallness of bounds on  $\delta c$
(see Table ). Using (\ref{103})-(\ref{photon}) we can write
\begin{equation} \delta c = c - \frac{n_0 c}{1 + \kappa} \end{equation}
or \begin{equation} \kappa = \frac{\delta c}{c - \delta c} =
\frac{\delta c}{c} + {\rm O}\left(\frac{\delta c^2}{c^2}\right)
\end{equation} where we have again used $n_0 \approx 1$. Bounds on
$\delta c/c$ can then be translated into bounds on $\kappa$.


\begin{table}[t]
\centerline{
\begin{tabular}{|c||c|c|c|c|c|c|}\hline
$\quad$ LIV tests $\quad$ & $\ \ \ \ \delta c /c\sim \ \ \ \ $ &\ \ \ \ Ref.\ \ \ \ \\
\hline \hline $\quad$Astrophysics $\quad$& $10^{-19}$ & \cite{lamor}
\\ \hline Atomic Physics   & $10^{-23}$  & \cite{hug} \\
\hline Laser Interferometry - Anisotropy & $10^{-16}$ & \cite{miche} \\
\hline Neutrino Sector & $10^{-19}$ & \cite{LSND}\\
\hline GZK cutoff (theoretical)& $10^{-20}$ & \cite{bertolami} \\
\hline Muon Collider (theoretical) & $ 10^{-21}$ & \cite{cole} \\
\hline Photon stability (theoretical) & $10^{-15}$ & \cite{cole}\\
\hline
\end{tabular}}
\caption{Bounds on $\delta c$ given by different Lorentz invariance
violation tests. \label{Table:lor}}
\end{table}

We  see that the smallness of the bounds implies, at very high
energy, that the chiral anomaly effects are effectively the same as
in the relativistic invariant case. Indeed, one has that  $ {(T)
_{LIV}}/ {(T)_R} -1$ takes values between $10^{-9} -10^{-22}$ and,
therefore, Lorentz invariance deviations are  experimentally almost
unattainable.

Therefore, the changes in the decay $\pi^0 \rightarrow 2 \gamma$ are
extremely small, which is consistent with the universal character of
the chiral anomaly.

One should note, however, that this conclusion could change
drastically if the other terms not considered in our analysis are
included. However, if we invoke the universal character of the
anomaly then  our assumption (\ref{true}) is reasonably justified.

\section{The chiral anomaly and the index theorem} \label{ap}

Let us consider an Hermitian Dirac operator that includes a minimal
Lorentz symmetry violation as discussed in the previous Section,
\begin{equation}\label{01}
    D_{m} = D+m\,, \quad {\rm with}\quad D = \Gamma^{\mu} D_{\mu}\,,
\end{equation}
where
\begin{equation}\label{02}
    D_\mu= i \frac{\partial}{\partial x^{\mu}} + A_{\mu}
\end{equation}
is the usual covariant derivative and the matrices $\Gamma^{\mu}$
are given in Eqs.\ (\ref{true}) and (\ref{omega}). A small mass $m$
has been introduced in order to deal with zero modes. At the end of
the calculation one should take the $m \to 0$ limit in a consistent
way \cite{GSMS}.

\medskip

 The $\Gamma$-matrices, depending on the constant tensor
$c^{\mu}_{\ \nu}$ and the parameter $Q$ which breaks Lorentz
symmetry, satisfy the {Clifford algebra} in Eq.\ (\ref{cli})  and
anticommute with $\gamma_5$  as in Eq.\ (\ref{07}). Their
(anti)commutator is given in Eq.\ (\ref{G6}).

\medskip

In order to to analyze the issue of chiral symmetry we follow the
Noether method  starting from  a $U(1)$ local chiral transformation,
\begin{equation}\label{08}
    \psi(x)\rightarrow e^{i \alpha(x) \gamma_{5}}\psi(x)
    \,, \quad
    \bar{\psi}(x)\rightarrow \bar{\psi}(x) e^{i \alpha(x)
    \gamma_{5}}\,,
\end{equation}
The fermionic Lagrangian changes as
\begin{equation}\label{09}
    \begin{array}{c}
      \mathcal{L}=\bar{\psi}(x) D_{m}\psi(x) \rightarrow
    \bar{\psi}(x) e^{i \alpha(x)\gamma_{5}} D_{m}
    e^{i \alpha(x) \gamma_{5}}\psi(x)=
     \\ \\
      =\mathcal{L} +\bar{\psi}(x) \left\{i \alpha(x)\gamma_{5}, D_{m}\right\}
    \psi(x) + O(\alpha^{2})\,.
    \end{array}
\end{equation}
The first order in $\alpha$ on the right hand side reduces, up to a
total divergence, to
\begin{equation}\label{010}
    \begin{array}{c}
      \delta\mathcal{L}=\bar{\psi}(x) \left\{i \alpha(x)\gamma_{5}, D_{m}\right\}
    \psi(x) = \\ \\
      =\alpha(x) \left( \partial_{\mu} \mathcal{J}^{\mu}_{5}(x)
    + 2im  \bar{\psi}(x) \gamma_{5}    \psi(x) \right)\,,
    \end{array}
\end{equation}
where the \emph{axial current} is now given in Eq.\ (\ref{j5}).

Let us now consider the functional integral
\begin{eqnarray}\label{013}
Z&=& \int \mathcal{D} \bar{\psi}\, \mathcal{D} \psi \,
    e^{-\int d^{4}x \, \mathcal{L} } \nonumber \\
&=& J[\alpha] \int \mathcal{D} \bar{\psi}\, \mathcal{D} \psi \,
    e^{-\int d^{4}x \mathcal{L} }
    \left(1+\int d^{4}x \, \delta \mathcal{L} +
    O(\alpha^{2})\right)\,, \nonumber\\
\end{eqnarray}
where
\begin{eqnarray}\label{014}
 J[\alpha] &=&  \frac{{\rm Det}\left(e^{i \alpha(x)\gamma_{5}}
D_{m}
    e^{i \alpha(x) \gamma_{5}}\right)}{{\rm Det}\, D_{m}}\nonumber \\
 &=&  1 - \int d^{4}x \, \alpha(x)
 \left\langle\partial_{\mu} \mathcal{J}^{\mu}_{5}(x)
   \right\rangle\nonumber\\
   &-& 2im \!\! \int\!\! d^{4}x\, \alpha(x)
   \left\langle\bar{\psi}(x) \gamma_{5}   \psi(x)
   \right\rangle
    +
     O(\alpha^{2})
     \label{noe}
\end{eqnarray}
is the Jacobian \cite{fuji,q-gamboa} arising from the change in the
fermionic measure under rotation (\ref{08}).

Being the Jacobian ill-defined (the Dirac operator eigenvalues grow
with no bound), one should introduce an appropriate regularization
($R$). We use a heat-kernel regularization so that, up to
$O(\alpha^{2})$
 terms, we have
\begin{eqnarray}
\log J[\alpha]\!\!\! &=&\!\!\!\left. ({\rm Tr}\log \left(D_{m} +
\left\{
    i \alpha(x)\gamma_{5}, D_{m}\right\}\right)-{\rm Tr}\log D_{m})
    \right\vert_{R}
    \nonumber
    \\  \nonumber
     \\
 \!\!\! &=&\!\!  \left.2 i {\rm Tr}\left( \alpha(x)\gamma_{5}
 \right)\right\vert_R \nonumber\\
 &=&
     \!\!   \left. 2 i \lim_{\Lambda \rightarrow\infty}
      {\rm Tr}
      \left( \alpha(x)\gamma_{5}
     e^{-\frac{{D_{m}}^{2}}{\Lambda^{2}}}
     \right)\right\vert_R \nonumber
      \\
       \nonumber
      \\
      &=& \!\!  2 i\!\!\! \lim_{\Lambda \rightarrow\infty}\!\!\!\int\!\!\! d^{4}x
      \!\!\!\int\!\!\! \frac{d^{4}k}{(2\pi)^{4}}\, {\rm tr}
      ( \alpha(x)\gamma_{5}\, e^{-i k \cdot x}
      e^{-\frac{{D_{m}}^{2}}{\Lambda^{2}}} \, e^{i k \cdot x} )
     \,,\nonumber\\
      \label{015}
\end{eqnarray}
 where
\begin{equation}\label{Dm2}
    {D_m}^2 = \mathbf{1}_4 M^{\mu\nu}D_{\mu}D_{\nu}
    +\frac{i}{4}\left[\Gamma^{\mu}, \Gamma^{\nu}\right]
    F_{\mu\nu} + O(m)\,.
\end{equation}

{A straightforward calculation taking into account that
\begin{equation}\label{016}
    {\rm tr}\left\{
    \gamma_{5}\left[\Gamma^{\mu},\Gamma^{\nu}  \right] \right\}
    =    \Omega^{\mu}_{\ \alpha} \Omega^{\nu}_{\ \beta } {\rm tr} \left\{
    \gamma_{5}
    \left[  \gamma^{\alpha} ,  \gamma^{\beta} \right]\right\}=0
\end{equation}
and ${\rm tr}\left\{  \gamma_{5} \right\}=0$ }leads to
\begin{eqnarray}
  \log J[\alpha]&=&     \displaystyle{
      - \frac{i}{16} \int d^{4}x
      \int\frac{d^{4}k}{(2\pi)^{4}}\, e^{-M^{\mu\nu} k_{\mu}
      k_{\nu}}} \nonumber
      \\  \nonumber \\
      &\times&\displaystyle{
      \, {\rm tr}
      \left( \alpha(x)\gamma_{5}
      \left[\Gamma^{\mu} , \Gamma^{\nu}\right]
      \left[\Gamma^{\alpha} , \Gamma^{\beta}\right] F_{\mu\nu}
      F_{\alpha\beta} \right)
      }
      \nonumber \\  \nonumber \\
      &=& \displaystyle{
      - \frac{i}{(16\pi)^{2}}\,
      {\left({\rm det}\, M \right)^{-1/2}}
      \Omega^{\mu}_{\ \rho} \,
    \Omega^{\nu}_{\ \sigma} \,
    \Omega^{\alpha}_{\ \kappa} \,
    \Omega^{\beta}_{\ \Omega}
      }  \nonumber
      \\ \nonumber \\
      &\times&
      \int \!\! d^{4}x
       {\rm tr}\,
      ( \alpha(x)\gamma_{5}
    \left[\gamma^{\rho} ,  \gamma^{\sigma}\right]
    \left[\gamma^{\kappa} ,  \gamma^{\Omega}\right]
     F_{\mu\nu}
      F_{\alpha\beta} ) , \nonumber\\
      \label{017}
\end{eqnarray}
up to $O(m)$ terms.

Using
\begin{equation}\label{018}
   {\rm tr}\left\{ \gamma_{5}
   \gamma^{\rho}\gamma^{\sigma}
   \gamma^{\kappa}\gamma^{\Omega}\right\}
 = 4 \, \epsilon^{\rho\sigma\kappa\Omega}
\end{equation}
one gets
\begin{eqnarray}
      \log J[\alpha]&=&
       \displaystyle{
      - \frac{i}{16 \pi^{2}} \int d^{4}x \,
      \, {\rm tr}
      \left( \alpha(x)\, \epsilon^{\rho\sigma\kappa\Omega}
     F_{\mu\nu}
      F_{\alpha\beta} \right) }\nonumber\\
      \nonumber \\   &\times& {\left({\rm det}\, M \right)^{-1/2}}
       \Omega^{\mu}_{\ \rho} \,
    \Omega^{\nu}_{\ \sigma} \,
    \Omega^{\alpha}_{\ \kappa} \,
    \Omega^{\beta}_{\ \Omega} \nonumber
    \\ \nonumber \\
    &=& \displaystyle{
      - \frac{i}{16 \pi^{2}} \int d^{4}x \,
      \, {\rm tr}
      \left( \alpha(x)\, \epsilon^{\mu\nu\alpha\beta}
     F_{\mu\nu}
      F_{\alpha\beta} \right)
      } \nonumber\\
      \nonumber \\& \times& {\left({\rm det}\, M \right)^{-1/2}}
      \, {\rm det}
      \left(\Omega  \right)
    \,.
\label{019}
\end{eqnarray}
Finally, taking into account Eq.\ (\ref{metric})
 one can see that all dependence on $\Omega^{\rho}_{\ \sigma}$
cancels out in the Jacobian,
\begin{equation}
 \log J[\alpha]=
\displaystyle{
 - \frac{i}{16 \pi^{2}}     \int d^{4}x \,
     \,\alpha (x) \,F_{\mu \nu}^{a} \, F_{\alpha
      \beta}^{b} \,
     \epsilon^{\mu\nu\alpha\beta}
     \ {\rm tr}\left(T_{a}T_{b}\right)
    }
\end{equation}
so that, after use of Eq.\ (\ref{noe}), one can write
\begin{eqnarray}
  \frac{i}{16 \pi^{2}} {\rm tr} \int d^{4}x &&\!\!\!\! \!\!\!\!\!\!\!\!\,
     \,\alpha (x) F_{\mu \nu}  \, F_{\alpha
      \beta}   \,
     \epsilon^{\mu\nu\alpha\beta}
     \
  \nonumber  \\
  &=&
  \int d^{4}x \, \alpha (x)\,  \langle \partial_{\mu}
       \mathcal{J}^{\mu}_{5}(x)\rangle
       \nonumber\\
       \nonumber \\
      &+ & \displaystyle{
     \lim_{m\rightarrow 0}
    2im \int d^{4}x \, \alpha (x)\,  \langle\bar{\psi}(x) \gamma_{5}
    \psi(x)\,
   \rangle} .\nonumber\\
\label{020}
\end{eqnarray}

Differentiating with respect to $\alpha$  one obtains the $U(1)$
anomaly equation in the form
\begin{eqnarray}
 \langle\,
     \partial_{\mu} \mathcal{J}^{\mu}_{5}(x)\rangle
     \!\!\!&+& \!\!\! \lim_{m\rightarrow 0}
    2im   \langle\bar{\psi}(x) \gamma_{5}   \psi(x)
 \rangle \nonumber \\ \nonumber\\
   &&  = \displaystyle{
      \frac{i}{16 \pi^{2}} \,{\rm tr}\left(F_{\mu \nu}(x)\, F_{\alpha
      \beta} (x)\right)
     \epsilon^{\mu\nu\alpha\beta}
    }\,.
\label{021}
\end{eqnarray}
Notice that the right hand side of this equation is insensitive to
the Lorentz symmetry breaking introduced by the tensor $c^{\mu}_{\
\nu}$ and the parameter $Q$.

\bigskip

Let us now consider the contribution of the zero modes. The mean
value of $ \bar{\psi}(x) \gamma_{5}   \psi(x) $ is given by
\begin{eqnarray}
\!\!\!\!\!\! \langle  \bar{\psi}(x)&&\!\!\!
\!\!\!\!\!\!\!\!\!\gamma_{5}
    \psi(x)  \rangle  \nonumber\\
    &=&\!\!
\frac{1}{Z}
 \displaystyle{
      \int\!\!  \mathcal{D}\bar{\psi} \mathcal{D}\psi \,
    e^{-\int \bar{\psi}(x) D_{m} \psi(x) \, d^{4}x} \,
    \bar{\psi}(x) \gamma_{5}  \psi(x) .} \nonumber\\
 \label{022}
\end{eqnarray}

Let $\varphi_{n}$ be the eigenvectors of $D$,
\begin{equation}\label{023}
    D \varphi_{n} = \Omega_{n} \varphi_{n}
    \quad \Rightarrow \quad
    D_{m} \varphi_{n} = (\Omega_{n}+m) \varphi_{n}\,,
\end{equation}
with
\begin{equation}\label{norm}
    \int {\varphi_{n}(x)}^{\dagger} \varphi_{m}(x) \, d^{4}x
    = \delta_{n,m}\,.
\end{equation}
Some of them can be zero modes of $D$. Since $ \left\{ \gamma_{5},
\Gamma^{\mu}\right\}=0$, one can always choose these zero modes with
a definite chirality,
\begin{equation}\label{024}
    D \varphi_{0,k}^{\pm} = 0
    \quad \Rightarrow \quad
    D_{m} \varphi_{0,k}^{\pm} = m \varphi_{0,k}^{\pm}\,, \ k=1,2,\dots
    n_{\pm}\,,
\end{equation}
with
\begin{equation}\label{025}
    \gamma_{5}\varphi_{0,k}^{\pm}= \pm \varphi_{0,k}^{\pm}\,.
\end{equation}

The integration variables in the functional integral can be expanded
as
\begin{equation}\label{026}
    \bar{\psi} = \sum_{n} \bar{c}_{n} {\varphi_{n}}^{\dagger}\,,
    \quad \psi =\sum_{n} {c}_{n} {\varphi_{n}}\,,
\end{equation}
and the integration measure be written as
\begin{equation}\label{027}
   \mathcal{D}\bar{\psi} \mathcal{D}\psi  = \prod_{n} d\bar{c}_{n}\,
   d{c}_{n}\,.
\end{equation}
Therefore,
\begin{eqnarray}
&&   \!\!\!\!\!\!\!\!\!\!    \langle\, \bar{\psi}(x) \gamma_{5}
    \psi(x) \,  \rangle  =\nonumber
    \\  \nonumber\\
 &&\!\!\!\!   \displaystyle{
    \!\!  \frac{1}{Z}\!\!  \int\!\!  \prod_{n} d\bar{c}_{n} \,d{c}_{n}
    e^{-\sum_{n}(\Omega_{n}+m) \bar{c}_{n} {c}_{n}}
    }
    \displaystyle{
    \sum_{p,q} \bar{c}_p
     {\varphi_{p}(x)}^{\dagger} \gamma_{5}  {\varphi_{q}(x)}c_q
     }
   \nonumber  \\ \nonumber \\
     &&
     \displaystyle{
     =\sum_{p,q}
     {\varphi_{p}(x)}^{\dagger} \gamma_{5} {\varphi_{q}(x)}
     \, \frac{1}{Z} \prod_n (\Omega_{n}+m)\, \frac{\delta_{p,q}}{(\Omega_{p}+m)}
     } \nonumber \\ \nonumber \\
   &&  \displaystyle{
     =\sum_{p}
     {\varphi_{p}(x)}^{\dagger} \gamma_{5}  {\varphi_{p}(x)}
     \, \frac{1}{(\Omega_{p}+m)}
     }\,,
\label{028}
\end{eqnarray}
since \begin{equation} Z=\prod_n (\Omega_{n}+m).
\end{equation}
Consequently,
\begin{eqnarray}
&& \!\!\!\!\!\! \displaystyle{\lim_{m \rightarrow 0} 2 i m
    \Big\langle\, \bar{\psi}(x) \gamma_{5}
    \psi(x) \, \Big\rangle}
    \displaystyle{
    =2 i \sum_{\Omega_{k}=0}
    {\varphi_{k}(x)}^{\dagger} \gamma_{5} {\varphi_{k}(x)}}
\nonumber      \\
\nonumber   \\
  &&  =2 i \left(\sum_{k=1}^{n_+}
    {\varphi_{0,k}^{+}(x)}^{\dagger}
     \varphi_{0,k}^{+}(x)
    -\sum_{k=1}^{n_-}
    {\varphi_{0,k}^{-}(x)}^{\dagger}
    \varphi_{0,k}^{-}(x)
    \right)\,. \nonumber\\
\label{029}
\end{eqnarray}
where $n_+$ ($n_-$) is the number of positive (negative) chirality
zero modes.

Now, with this result one can integrate over   space on both sides
of Eq.\ (\ref{021}) to obtain
\begin{eqnarray}
  \int d^{4}x \sum_{k=1}^{n_+}
    {\varphi_{0,k}^{+}(x)}^{\dagger}  {\varphi_{0,k}^{+}(x)}
 -\int d^4x
    \sum_{k=1}^{n_-}
    {\varphi_{0,k}^{-}(x)}^{\dagger} {\varphi_{0,k}^{-}(x)}
\nonumber    \\ \nonumber  \\
    \displaystyle{=\frac{1}{32 \pi^{2}} \int d^{4}x \,{\rm tr}
    \left\{F_{\mu \nu}^{b}(x)\, F_{\alpha
      \beta}^{c}(x)\right\}\,
     \epsilon^{\mu\nu\alpha\beta}
     \   }\,, \nonumber\\
     \label{030}
\end{eqnarray}
where we have discarded the contribution of the total divergence of
$\mathcal{J}^{\mu}_{5}$.  Eq.(\ref{030}) can be written in the form
\begin{equation}
 n_+  - n_-
  =\frac{1}{32 \pi^{2}} \int d^{4}x  {\rm tr}\left\{
   F_{\mu \nu}(x)\, F_{\alpha
      \beta}(x)\right\}
     \epsilon^{\mu\nu\alpha\beta}\,,
     \label{031}
\end{equation}
which is nothing but the \emph{index theorem} for the Dirac operator
$D$.

Then, neither the anomaly (Eq.\ (\ref{021})) nor the index theorem
(Eq.\ (\ref{031})) are affected by this minimal Lorentz symmetry
violation. Only the expression of the axial current
$\mathcal{J}^{\mu}_{5}$ is changed as in Eq.\ (\ref{j5}).

Let us end this section by noting that an investigation on the
relation between Lorentz violation and vector models with a
Wess-Zumino term which can be connected with models containing
chiral fermions has been reported in \cite{AnSol}. In that case a
dynamical Lorentz violation is described as the nonperturbative
counterpart of perturbative unitarity breaking in chiral gauge
theories due to gauge anomalies.

\section{Summary and discussion}

In this work we explored the fate of the chiral anomaly in a
fermionic model in which the Lorentz symmetry is explicitly broken
by terms which  preserve chiral symmetry at the classical level
and the form of the Clifford algebra satisfied by the $\Gamma^\mu$
matrices replacing Dirac matrices in the Lorentz invariance violating
fermionic Lagrangian.

 On rotational invariance grounds, only the $c^0_{\ 0}$
component was taken as non-vanishing in Eq.\ (\ref{100}),  so that
the energy-momentum relation is changed in the sense that each
particle has a maximum attainable velocity (see eqs.\
(\ref{102})-(\ref{103})) which depends only on the dimensionless
parameter $\kappa$ (and is independent of the parameter $Q$).

Now, in view of the connection, through the Dirac operator index,
between the anomaly and a topological object (the Chern-Pontryagin
index) one should expect that the anomaly itself as well as its
physical implications (like those related to the $\pi^0 \to 2\gamma$
decay) remain unaltered.

Concerning the anomaly, we have shown, within the path-integral
approach and using a heat-kernel regularization, that the Fujikawa
Jacobian is not modified. This result was obtained by regularizing 
the path integral measure with the
same operator that plays the role of the Dirac operator in the
classical action, namely that with a minimal Lorentz violation that
 classically preserves chiral invariance.

Moreover, the Noether method yields to an anomaly equation for the
chiral current that is formally the same as in the Lorentz invariant
case, except that the divergence term contains the modified axial
current (\ref{j5}). However, since the contribution of such term
vanishes when integrated over all space, the index theorem equation
remains unaltered.

We have also discussed  within a Lorentz violating framework, the
issue of the $\pi^0 \to 2\gamma$ decay, which is connected to the
chiral anomaly.

Since the adopted Lorentz symmetry breaking implies different
velocities for different massless particles, fermion velocities in
the triangle diagram differ from the photon one (Eqs.\
(\ref{103}-\ref{photon})). Then, the $\pi^0$ decay amplitude in the
chiral limit is modified and the change is proportional to $\kappa$
(Eq.\ (\ref{esti})), a parameter controlling Lorentz violation. Now,
different experimental and phenomenological tests show the smallness
of $\kappa$ so that changes in the $\pi^0 \to 2\gamma$ cannot be
detected. However if other effects such as addition of CPT violating
terms were considered, the form of the chiral anomaly could be
affected and experimental consequences in processes as that of the
$\pi^0$ decay could become detectable.

\vspace{0.3 cm}

\noindent\underline{Acknowledgements}:  We would like to thank to
J.~Alfaro, M. Asorey, H.~O.~Girotti and L.~Alvarez-Gaum\'e for
passionate discussions on the subject during the CEFIMAS Buenos
Aires Workshop on May 2007. We would like to thank also Professors W. Bietenholz,
R. Jackiw and V. A. Kostelecky by useful comments on this
manuscript.

This work was partially supported by FONDECYT-Chile and CONICYT
grants 1050114, 1060079 and 21050196, PIP6160-CONICET,
PIC-CNRS/CONICET, BID 1728OC/AR PICT20204-ANPCYT grants and by CIC
and UNLP (11/X381 and 11/X450), Argentina.

\newpage


\begin{thebibliography}{}
\bibitem{weinberg} S. Weinberg, {\it Quantum Field Theory}, Vol. I and II, Cambridge University
Press (1995).
\bibitem{CPT} G. Amelino-Camelia, J. Ellis, N. Mavromatos, D. V. Nanopoulos
and S. Sarkar, {\it Nature}, {\bf 393}, 763 (1998); D. Sudarsky, L.
Urrutia and H. Vucetich, {\it Phys. Rev.} {\bf D68}, 024010 (2003);
J. Alfaro, H. Morales-Tecotl and L. F. Urrutia, {\it Phys Rev. Lett}
{\bf 84}, 2318 (2000); {\it ibid}, {\it Phys. Rev.} {\bf D65},
103502 (2002); N.R. Bruno, G. Amelino-Camelia and J.
Kowalski-Glikman, {\it Phys. Lett.} {\bf B522}, 133 (2001); G.
Amelino-Camelia,  {\it Int. J. Mod. Phys.} {\bf D11}, 35
(2002);Yi-Fu Cai, and Yun-Song Piao, gr-qc/0701114.
\bibitem{kost}V. A. Kosteleck\'y and S. Samuel, {\it Phys. Rev.} {\bf D39}, 683
(1989); V. A. Kosteleck\'y {\it Phys. Rev.} {\bf D69},105009 (2004);
V. A. Kosteleck\'y and R. Potting, {\it Phys. Rev.} {\bf D51}, 3923
(1995); D. Colladay and V.A. Kosteleck\'y,  Phys. Lett. {\bf
  B511}, 209 (2001); V. A. Kosteleck\'y, R. Lehnert, {\it Phys. Rev.}  {\bf D63}, 065008 (2001);
R. Bluhm and  V. A. Kosteleck\'y, {\it Phys. Rev. Lett}. {\bf 84},
1381 (2000); V. A. Kosteleck\'y and  C. D. Lane, {\it Phys. Rev}.
{\bf D60}, 116010 (1999); R. Jackiw and V. A. Kosteleck\'y, {\it
Phys. Rev. Lett}. {\bf 82}, 3572 (1999); D. Colladay, V. A.
Kosteleck\'y, {\it Phys. Rev}.  {\bf D58}, 116002 (1998); O.
Bertolami, D. Colladay, V. A. Kostelecky, R. Potting, {\it Phys.
Lett.} {\bf B395}, 178 (1997).
\bibitem{jackiw} R. Jackiw and S.Y. Pi, {\it{Phys. Rev}}. {\bf D68}, 104012
(2003); Z. Guralnik,R. Jackiw, S.Y. Pi, A.P. Polychronakos, {\it
Phys. Lett.} {\bf B517}, 450 (2001);S. Carroll, R. Jackiw and G.
Field, {\it Phys. Rev.} {\bf D41}, 1231 (1990); R. Jackiw and S. Y.
Pi, Phys.Rev.D68:104012,2003
\bibitem{nos} J.M. Carmona, J.L. Cortes, J. Gamboa, and F. Mendez, {\it JHEP} {\bf 0303}, 058 (
2003), {\it ibid} {\it Phys. Lett.} {\bf B565}, 222 (2003);  A. Das,
J. Gamboa, F.  Mendez, J. Lopez-Sarrion, {\it JHEP}  {\bf 0405}, 022
(2004);  J. Carmona, J. L. Cortes, A. Das, J. Gamboa, F. Mendez,
{\it  Mod. Phys. Lett.} {\bf A21}, 883 (2006);  J. Gamboa and J.
Lopez-Sarrion, {\it Phys. Rev.} {\bf D71}, 067702 (2005);  H.
Falomir, J. Gamboa, J. Lopez- Sarrion, F. Mendez, A.J. da Silva.
{\it Phys. Lett.} {\bf B632}, 740 (2006); {\it ibid}, {\it Phys.
Rev.} {\bf D74}, 047701 (2006); J. Gamboa, J. Lopez-Sarrion, A. P.
Polychronakos,  {\it Phys. Lett.} {\bf B634}, 471 (2006); P. Arias,
A. Das, J. Gamboa, J. Lopez-Sarrion and F. Mendez, hep-ph/0608007,
{\it Phys. Lett.} {\bf B} (2007), in press; A. Das, J. Gamboa, J.
Lopez- Sarrion, F. A. Schaposnik, {\it Phys. Rev.} {\bf D72}, 107702
(2005); M. Gomes, J.R. Nascimento, E. Passos, A.Yu. Petrov, A.J.da
Silva, hep-th/07041104; A.F. Ferrari, M. Gomes, J.R. Nascimento, E.
Passos, A.Yu. Petrov, A.J. da Silva, hep-th/0609222.
\bibitem{jackiw1}  A. A.
Andrianov, P. Giacconi, R. Soldati, {\it JHEP} {\bf  0202}, 030
(2002); J. Alfaro, A. A. Andrianov, M. Cambiaso, P. Giacconi, R.
Soldati, {\it Phys. Lett.} {\bf B639}, 586 (2006).
\bibitem{bertolami} O. Bertolami, and C.S. Carvalho, {\it Phys.Rev.} {\bf D61}, 103002
(2000).
\bibitem{mewes} V. A. Kosteleck\'y and M. Mewes, {\it Phys.Rev.} {\bf
D69},016005 (2004).
\bibitem{ko}   V. A. Kosteleck\'y and M. Mewes, {\it Phys. Rev.} {\bf D66}, 056005 (2002).
\bibitem{cole1}S. Coleman and S. L. Glashow, {\it Phys.Lett} {\bf B405}, 249 (1997)
\bibitem{cole} S. Coleman and S. L. Glashow, {\it Phys. Rev.} {\bf D59}, 116008 (1999)
\bibitem{GG} M.~C.~Gonzalez-Garcia and M.~Maltoni,
  arXiv:0704.1800 [hep-ph].
\bibitem{adler} S. L. Adler, {\it Phys. Rev.} {\bf 177}, 2426 (1969).
\bibitem{jackiw2} J. S. Bell and R. Jackiw, {\it Nuo. Cim.} {\bf 60A}, 47 (1969).

\bibitem{lamor} S.K. Lamoreaux, J.P. Jacobs, B.R. Heckel, F.J. Raab
and E.N. Fortson, {\it Phys. Rev. Lett} {\bf 57}, 3125 (1986); B.
Altschul, {\it Phys. Rev.} {\bf D75},041301 (2007); {\it ibid}, {\it
Phys. Rev. Lett.} {\bf 96}, 201101 (2006)
\bibitem{GSMS}
  R.~E.~Gamboa Saravi, M.~A.~Muschietti and J.~E. Solo\-min,
  {\it Commun.\ Math.\ Phys. }  {\bf 93}, 407 (1984).

\bibitem{fuji} K. Fujikawa, {\it Phys. Rev. Lett.} {\bf 42}, 1195 (1979); {\it ibid}, {\it Phys. Rev.} {\bf
D21}, 2848 (1980).
\bibitem{q-gamboa} R.E.\ Gamboa Sarav{\'\i}, M.E.\ Muschietti, F. Schaposnik and J.\ Solomin, {\it
Annals Phys}., {\bf157}, 360 (1984).
\bibitem{miche} P.L. Stanwix, M.E. Tobar, P. Wolf, C.R. Locke and
E.N. Ivanov, {\it Phys. Rev.} {\bf D74}, 081101 (2006); S. Herrmann,
A. Senger, E. Kovalchuk, H. M\"{u}ller and A. Peters, {Phys. Rev. Lett.}
{\bf 95}, 150401 (2005); P. Antonini, M. Okhapkin, E. Goeklue and S.
Schiller, {\it Phys. Rev.} {\bf A71}, 050101 (2005); J. A. Lipa, J.
A. Nissen, S. Wang, D. A. Stricker and D. Avaloff, {\it Phys. Rev.
Lett.} {\bf 90}, 060403 (2003).
\bibitem{hug} P.Wolf, F. Chapelet, S. Bize and A. Clairon, {\it Phys. Rev.
Lett.} {\bf 96}, 060801 (2006)
\bibitem{LSND} L.B. Auerbach {\it et al.}, {\it Phys. Rev.} {\bf
D72}, 076004 (2005); T. Katori {\it et al. }, {\it Phys. Rev.} {\bf
D74}, 105009 (2006).
\bibitem{AnSol}
  A.~A.~Andrianov and R.~Soldati,
  vector field models with Wess-Zumino
  Phys.\ Rev.\  D {\bf 51} (1995) 5961.
\end{thebibliography}
\end{document}